\documentclass[aps,prb,reprint]{revtex4-1}
\usepackage{amsmath, amssymb}
\usepackage{graphicx}
\usepackage[usenames,dvipsnames]{color}

\newcommand{\ket}[1]{{| {#1}  \rangle}}
\newcommand{\bra}[1]{{\langle {#1} |}}
\newcommand{\expval}[1]{{\langle {#1} \rangle}}
\newcommand{\braket}[2]{{\langle {#1} | {#2} \rangle}}
\DeclareMathOperator{\tr}{tr}

\begin{document}

\title{Spatial noise filtering through error correction for quantum sensing}

\author{David Layden}
\author{Paola Cappellaro}

\affiliation{Research Laboratory of Electronics and Department of Nuclear Science and Engineering, Massachusetts Institute of Technology, Cambridge, Massachusetts 02139, USA}

\begin{abstract}
Quantum systems can be used to measure various quantities in their environment with high precision. Often, however, their sensitivity is limited by the decohering effects of this same environment. Dynamical decoupling schemes are widely used to filter environmental noise from signals, but their performance is limited by the spectral properties of the signal and noise at hand. Quantum error correction schemes have therefore emerged as a complementary technique without the same limitations. To date, however, they have failed to correct the dominant noise type in many quantum sensors, which couples to each qubit in a sensor in the same way as the signal. Here we show  how quantum error correction can correct for such noise, which dynamical decoupling can only partially address. Whereas dynamical decoupling exploits temporal noise correlations in signal and noise, our scheme exploits spatial correlations. We give explicit examples in small quantum devices and demonstrate a method by which error-correcting codes can be tailored to their noise. 
\end{abstract}

\maketitle

\section*{Introduction}

Quantum sensors exploit the strong sensitivity of quantum systems to external disturbances. They typically measure an external quantity---a DC signal in the most common scenario---whose value is manifest as a parameter in the sensor's Hamiltonian. This parameter can then be estimated by preparing the sensor in a superposition of two energy eigenstates and letting them acquire a relative phase that depends on the unknown quantity, which is then read out \cite{degen:2017}.

This strong sensitivity of quantum systems to their environment is a double-edged sword, however, as it also brings about decoherence. For quantum sensors, decoherence means that not only the relative phase---but also the phase uncertainty---between energy eigenstates grows with time. The competing contributions of phase (signal) and phase uncertainty (noise) both stemming from a sensor's environment fundamentally limit the precision with which it can measure external quantities. A central way to improve the sensitivity of a quantum sensor is therefore to operate it in a way that filters out environmental disturbances which constitute noise, leaving those which constitute signal.

Dynamical decoupling (DD) sequences provide a family of such filters, which are ubiquitous in quantum sensing experiments\cite{viola:1998, ban:1998, degen:2017}. They comprise a series of control pulses (or continuous controls) applied to a sensor, which modify its response to perturbations of different frequencies. This allows them to filter noise from signal in a quantum sensor on the basis of frequency; typically, by forming an effective narrow-band filter around an AC signal, thus letting the signal imprint on the sensor while suppressing much of the noise outside the main passband. Since DD sequences act as filters in the frequency domain, their main limitations for quantum sensing have to do with the spectra of signal and noise: First, DD-based sensors can only measure a narrow frequency band of the signal at a time. Moreover, they cannot directly measure DC signals, as DD suppresses both noise and signal at low frequencies. Second, they remain vulnerable to high-frequency noise components, which limit the sensitivity they can achieve\cite{biercuk:2011}. A complementary family of filters based instead on quantum error correction has recently emerged, which do not suppress noise on the basis of frequency, and therefore do not share these same limitations\cite{arrad:2014, kessler:2014, ozeri:2013, dur:2014}.

The canonical scheme for error-corrected quantum sensing (ECQS), illustrated here with a Lindblad description, is: (i) to prepare a superposition of logical energy eigenstates, (ii) to let the sensor evolve for a time $\Delta t$ under the Liouvillian $\mathcal{L} = -i \mathcal{H} + \mathcal{D}$, where the Hamiltonian superoperator $\mathcal{H}(\rho) = [H, \rho]$ is proportional to the parameter one wants to estimate (that is, the signal\cite{giovannetti:2006}), and (iii) to apply a recovery operation $\mathcal{R}$ which seeks to correct the effects of the noise from 
\begin{equation}
\mathcal{D}(\rho)= \sum_i L_i \rho L_i^\dagger - \frac{1}{2} \{ L_i^\dagger L_i, \rho \},
\label{eq:D_diag}
\end{equation}
where $\{ L_i \}$ are the Lindblad error operators describing decoherence, which can be interpreted as quantum jumps\cite{lidar:2013}. (The recovery is approximated as being instantaneous.) Steps (ii) and (iii) are repeated until (iv) the final state is read out after a total time $t$. In the limit where (ii)--(iii) are fast and repeated many times ($\Delta t \rightarrow 0$ with $t$ finite), the sensor evolves stroboscopically as
\begin{equation}
\mathcal{L}_\text{eff} = -i \mathcal{R H}
+ \mathcal{R D} + O \big (||\mathcal{L}|| \, \Delta t \big)
\label{eq:L_eff}
\end{equation}
according to Chernoff's theorem (ref.\ \onlinecite{chernoff1968} p.\ 241, see also ref.\ \onlinecite{layden:2016}). If $\mathcal{R D} = 0$ but $\mathcal{R H} \neq 0$ on logical states then ECQS can approach a noiseless sensing limit by making $\Delta t$ sufficiently short compared to the noise strength.

For ECQS to provide such a noiseless $\Delta t \rightarrow 0$ limit ($\mathcal{RD} |_\text{code} = 0$), the error operators for the sensor must satisfy the usual Knill-Laflamme condition \cite{knill:1997, beny:2011}:
\begin{equation}
P L_i^\dagger L_j P \propto P
\label{eq:KL}
\end{equation}
for all $i,j \ge 0$, where $P = P^\dagger$ projects onto the codespace and $L_0 := I$. For the signal to survive in this limit ($\mathcal{RH}|_\text{code} \neq 0$), however, $H$ must not be fully correctable:
\begin{equation}
P H P \not \propto P.
\label{eq:KL_H}
\end{equation}
Naturally then, $H$ must not be a linear combination of $L_i^\dagger L_j$ terms. Indeed, it was recently proven that there exists a code with projector $P$ satisfying these ECQS conditions, Eqs.~\eqref{eq:KL} and \eqref{eq:KL_H}, if and only if $H \notin \mathcal{S}$ where $\mathcal{S} = \text{span} \{ I, L_i, L_i^\dagger, L_i^\dagger L_j \}$ is the so-called Lindblad span\cite{demkowicz:2017, zhou:2017}.

An archetypal example of ECQS to measure a DC signal $\omega$ was proposed in refs.\ \onlinecite{ozeri:2013, dur:2014, reiter:2017}: it involves a three-qubit sensor with $H = \sum_{i=1}^3 \frac{\omega}{2} \, Z_i$ subject to independent bit-flip errors on each qubit ($L_i = X_i$). Initializing the sensor in $\ket{+_\textsc{l}} = \frac{1}{\sqrt{2}} (\ket{0_
\textsc{l}} + \ket{1_
\textsc{l}})$, where $\ket{0_
\textsc{l}} = \ket{000}$ and $\ket{1_
\textsc{l}} = \ket{111}$, the errors can be detected and corrected by the bit-flip code recovery $\mathcal{R}$ (ref.\ \onlinecite{oreshkov:2007}), while the signal $\omega$ imprints through $H$ as a relative phase in the encoded state. The signal is unimpeded by the frequent applications of $\mathcal{R}$ as it couples to the qubits through a different operator than the noise, giving $H \notin \mathcal{S}$, and hence $\mathcal{RD} = 0$ but $\mathcal{RH} \neq 0$ on logical states. Therefore, the scheme enables near-noiseless sensing of $\omega$ for short $\Delta t$. To our knowledge, all explicit ECQS schemes to date for multi-qudit sensors operate similarly, correcting only for noise which couples to the sensor via different operators than the signal \cite{arrad:2014, kessler:2014, ozeri:2013, dur:2014, herrera:2015, plenio:2016, gefen:2016, bergmann:2016, sekatski:2016, unden:2016, cohen:2016, reiter:2017, matsuzaki:2017}. If instead the jump operators were $L_i = Z_i$ (that is, if the noise coupled to each qubit in the same way as the signal), Eqs.\ \eqref{eq:KL} and \eqref{eq:KL_H} could not be satisfied, as no code could filter noise from signal since $H \in \mathcal{S}$.

This example illustrates a deficiency in ECQS schemes to date, of both practical and fundamental importance. In many quantum sensors, noise which couples through the same operators as the signal is the dominant source of decoherence (often by several orders of magnitude~\cite{biercuk:2009, witzel:2010, bluhm:2011, doherty:2013, muhonen:2014, orgiazzi:2016}), and thus imposes the main limit on achievable sensitivity. Schemes which do not protect against such a central noise source can offer only a limited advantage in practice. More fundamentally, the outsized importance of such noise derives from a tension at the core of quantum sensing: A good sensor must couple strongly to the quantity it is to measure. Such strong coupling, in turn, renders the device highly sensitive to fluctuations in this quantity, thus bounding the duration of coherent sensing. While DD can suppress certain frequency components of this noise, explicit ECQS schemes to date cannot address it at all, to our knowledge. Moreover, refs.~\onlinecite{demkowicz:2017, zhou:2017} make no pronouncements as to whether, or when, it is possible to correct for noise that couples to a sensor through the same operator in its Hamiltonian as the signal. This potential Achilles' heel would seem to largely determine the practical value of error correction in many quantum sensors.

We propose a general error-correction scheme to filter out noise which couples to each qubit in a quantum sensor identically to the signal. Our scheme is not frequency-selective, and therefore does not share the same limitations of dynamical decoupling. The key insight is that Eqs.\ \eqref{eq:KL} and \eqref{eq:KL_H} can be viewed not only as a condition on how signal and noise couple to the sensor, as with existing codes, but also as a condition on the spatial profiles of each in the sensor. Indeed, we show that quantum error correction can filter noise from signal on the basis of their respective spatial correlations, much like DD filters based on temporal correlations (i.e., frequency profiles). 

\section*{Results}

We first consider the task of measuring a DC signal $\omega_0$, inaccessible through DD, in the presence of background noise $\delta \omega$. We assume a generic sensor comprising $N$ qubits, each coupled locally to the external field $\omega(\vec{x},t) = \omega_0 + \delta \omega(\vec{x}, t)$ as
\begin{equation}
H(t) = \frac{1}{2} \sum_{i=1}^N \omega (\vec{x}_i, t) \, Z_i
\label{eq:H_stochastic}
\end{equation}
in a suitable reference frame, where $\vec{x}_j$ is the position of qubit $j$ and $\delta \omega$ describes zero-mean stationary fluctuations. We take $\delta \omega$ to be Gaussian white noise with strength $\sim \! 1/T_2$, for both mathematical convenience and to highlight the ability of error correction to filter high-frequency background noise beyond the reach of DD. We allow, however, for general spatial correlations between the fluctuations at positions $\vec{x}_i$ and $\vec{x}_j$:
\begin{equation}
\big \langle \delta \omega( \vec{x}_i, t) \,
\delta \omega( \vec{x}_j, 0) 
\big \rangle
= \frac{2  \delta(t)}{T_2} \, c_{ij}.
\end{equation}
Thus, any qubit prepared in the state $\ket{+}$ will precess about its $z$-axis at a rate $\omega_0$ and lose phase coherence as $|\expval{\sigma_+^{(j)}}| = \exp(- t/T_2)$ (ref.\ \onlinecite{kubo:1962}). The coefficients $c_{ij} \in [-1,1]$ describe the noise correlations between positions $\vec{x}_i$ and $\vec{x}_j$: The extreme values of $c_{ij}= \pm 1  $, for instance, describe identical ($+1$) or opposite ($-1$) fluctuations in $\omega$ at either position, whereas $c_{ij}=0$ means no correlation. Naturally $c_{ii}=1$. Note that while this noise is external to the sensing degrees of freedom, it may still arise from within the experimental device, e.g., from the surrounding nuclear bath in the case of spin qubits.

It is convenient to express the sensor's dynamics as a master equation $\dot{\rho} = \mathcal{L}(\rho)$, where
\begin{equation}
\mathcal{L}(\rho)
=
-i[H_0,  \rho]
+
\frac{1}{2 T_2}
\sum_{i,j=1}^N
c_{ij} 
\Big(
Z_i \rho Z_j
 - \frac{1}{2} \{ Z_i Z_j, \rho \}
 \Big)
 \label{eq:Lindblad}
\end{equation}
and $H_0 = \frac{\omega_0}{2} \sum_{i=1}^N Z_i$ (refs.\ \onlinecite{cheng:2004, cheng:2005}). Eq.~\eqref{eq:Lindblad} can be cast in the form of \eqref{eq:D_diag}, thus removing dissipative cross terms, by diagonalizing the correlation matrix $C = (c_{ij})_{i,j \ge 1}$ to yield operators $L_i = \sqrt{\lambda_i} \vec{v}_i \cdot \vec{Z}$. Here, $C \vec{v}_i = \lambda_i \vec{v}_i$ and $\vec{Z} = (Z_1, \dots, Z_N)$. These $L_i$'s can be interpreted as the sensor's quantum jumps, while the $Z_i$'s in \eqref{eq:Lindblad} cannot\cite{lidar:2013}. Crucially, the ECQS conditions, \eqref{eq:KL} and \eqref{eq:KL_H}, deal with the quantum jump operators $L_i$'s, not the bare $Z_i$'s. This distinction is critical because it opens the possibility of engineering a sensor such that $C$ has a vanishing eigenvalue $\lambda_k = 0$, thus suppressing $L_k$. Generically, the Lindblad span $\mathcal{S}$ will then fail to contain $H_0$ due to this ``missing" jump operator, opening the door for ECQS.

The requirement that $H_0 \notin \mathcal{S}$ for conditions \eqref{eq:KL} and \eqref{eq:KL_H} can be restated for signal and noise which couple identically to the sensor (in the sense of Eq.~\eqref{eq:H_stochastic}) as
\begin{equation}
\vec{h} \notin \text{col}(C),
\label{eq:hcolC}
\end{equation}
where $\vec{h} = (1, \dots, 1) \in \mathbb{R}^N$ so that $H_0 = \frac{\omega_0}{2} \, \vec{h} \cdot \vec{Z}$, and $\text{col}(C)$ is the column space of $C$. A full proof is given in the Methods; we note here simply that Eq.~\eqref{eq:hcolC} enforces two things: (i) that $\text{det}(C) = 0$ and thus $L_k = 0$ for some $k$, and (ii) that $H_0$ is not composed only of non-vanishing $L_i$'s. It ensures that the signal and noise can be fully distinguished by the recovery operation on the basis of their respective spatial profiles (a requirement we will later relax).

Consider for example $N=3$ sensing qubits positioned so that the fluctuations satisfy $c_{ij} = -\gamma/2$ for each pair $i\neq j$, where $\gamma \in [0,1]$ describes the noise correlation strength. (Specifically, $\gamma=0$ produces vanishing correlations whereas $\gamma=1$ gives the strongest correlations possible.) Notice that for $\gamma=1, C \vec{h} = \vec{0}$, and so $\vec{h} \notin \text{col}(C)$. The jump operators in the ECQS conditions are not $Z_1$, $Z_2$ and $Z_3$, but rather
\begin{align}
L_1 &= \frac{\sqrt{2+\gamma}}{2} (  Z_1- Z_3 ), \nonumber \\ 
L_2 &= \sqrt{\frac{2+\gamma}{12}}  ( Z_1 - 2 Z_2 + Z_3 ), \\
L_3 &= \sqrt{\frac{1-\gamma}{3}} (Z_1 + Z_2 + Z_3), \nonumber 
\end{align}
found by diagonalizing $C$. Observe that the global noise mode, $L_3$, becomes subdominant for larger values of $\gamma$, until it vanishes completely when $\gamma=1$, at which point $H_0 \notin \mathcal{S}$ as expected. Notice also that $P = \ket{0_\textsc{l}}\!\bra{0_\textsc{l}} + \ket{1_\textsc{l}} \! \bra{1_\textsc{l}}$ with logical states
\begin{align} \label{eq:ex1_states}
\ket{0_\textsc{l}} &= \frac{1}{\sqrt{3}} \big( \ket{100} + \ket{010} + \ket{001} \big) \\
\ket{1_\textsc{l}} &= \frac{1}{\sqrt{3}} \big( \ket{011} + \ket{101} + \ket{110} \big) \nonumber
\end{align} 
satisfies the conditions \eqref{eq:KL} and \eqref{eq:KL_H} when $\gamma \rightarrow 1$, despite the signal and noise both coupling to each sensing qubit identically. Contrast this with the usual phase-flip code, which corrects for $Z_1, Z_2, Z_3$ and all linear combinations thereof, including $H_0$ (ref.\ \onlinecite{lidar:2013}). Eq.\ \eqref{eq:ex1_states} instead defines a weakened version of the phase-flip code, which corrects for linear combinations of $L_1$ and $L_2$, but not for $\vec{v}_3 \cdot \vec{Z} \propto Z_1 + Z_2 + Z_3$. Accordingly, as $\gamma \rightarrow 1$ it can fully correct the noise ($\mathcal{RD} |_\text{code} = 0$) while allowing the signal to still imprint on the logical states ($\mathcal{RH} |_\text{code} \neq 0$). In particular, the dynamics at logical level for $\Delta t \rightarrow 0$ are generated by the effective Hamiltonian $H_\text{eff} = \frac{\omega_0}{2} Z_\textsc{l}$ and the effective Lindblad error operator 
\begin{equation}
L_\text{eff} = \sqrt{\frac{1-\gamma}{6 T_2} } Z_\textsc{l},
\end{equation}
where $Z_\textsc{l}$ is the logical $\sigma_z$ (ref. \onlinecite{zhou:2017}). Notice that $L_\text{eff} \rightarrow 0$ for $\gamma \rightarrow 1$, so the logical dynamics is less noisy for stronger correlations. A detailed calculation of the effective Liouvillian for this example---as well as for an example with positive noise correlations---is provided in the Supplementary Information.

\begin{figure}[h]
\raggedleft
\includegraphics[width=0.48\textwidth]{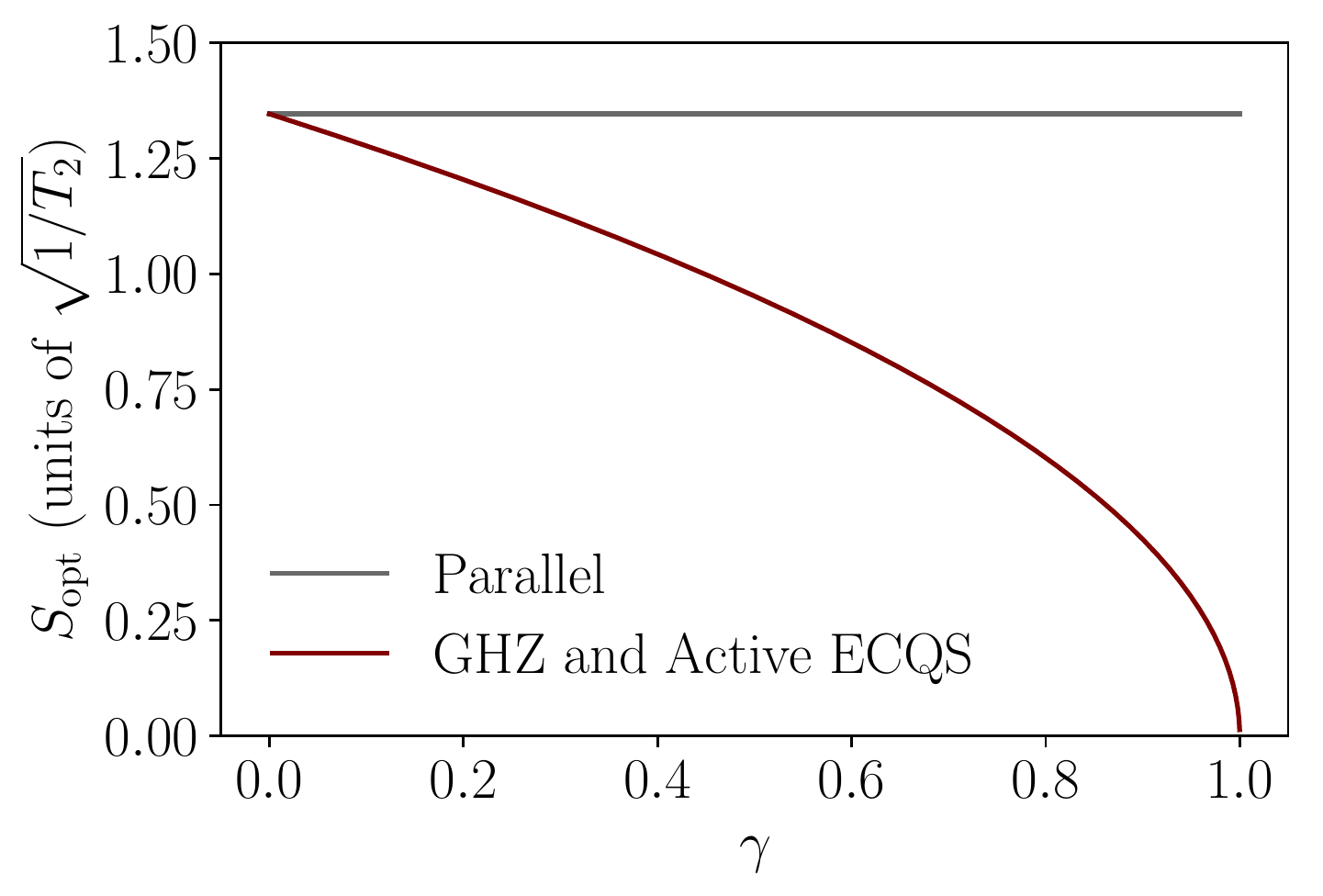}
\caption{The achievable sensitivity with different schemes for a 3-qubit sensor under noise correlations $c_{ij}=-\gamma/2$, for all $i\neq j$. Note that sensitivity is a measure of the smallest resolvable signal per unit time, so a smaller sensitivity indicates better performance. The active recovery and GHZ schemes use codewords \eqref{eq:ex1_states} and \eqref{eq:ex2_states} respectively, whereas the parallel scheme operates the qubits individually, without entanglement. The active and GHZ schemes give identical performance (in the regime of $\Delta t \rightarrow 0$ for the former), and both outperform parallel sensing by a factor that grows with the correlation strength $\gamma$. The details of this calculation are provided in the Supplementary Information.}
\label{fig:neg_sensitivity}
\end{figure}

Another possible code for this $C$ uses
\begin{equation}
\ket{0_\textsc{l}'} = \ket{000} \qquad \quad
\ket{1_\textsc{l}'} = \ket{111}.
\label{eq:ex2_states}
\end{equation}
When $\gamma \rightarrow 1$ it also satisfies the ECQS conditions, although its recovery procedure is trivial because $\text{span} \{ \ket{0_\textsc{l}'}, \ket{1_\textsc{l}'} \}$ is a decoherence-free subspace (DFS) within which $H_0$ acts non-trivially\cite{lidar:1998}. ECQS with this code is therefore a Greenberger-Horne-Zeilinger (GHZ) sensing scheme\cite{GHZ}. The performance of a quantum sensor can be quantified by the sensitivity it achieves; that is, the smallest signal it can detect per unit time\cite{degen:2017}. Sensitivity therefore also provides a way to benchmark ECQS schemes: a more effective scheme allows one to resolve a smaller signal per unit time, thus giving lower (i.e., better) sensitivity. The sensitivity offered by the codes in Eqs.~\eqref{eq:ex1_states} and \eqref{eq:ex2_states} is plotted in Fig.~\ref{fig:neg_sensitivity} as a function of $\gamma$. Both approach noiseless sensing when $\gamma \rightarrow 1$.

In general, a DFS is a code for which $\ket{0_\textsc{l}}$ and $\ket{1_\textsc{l}}$ are degenerate eigenvectors of all $L_i$. This means that $EP = \mu_E P$ for all $E \in \mathcal{S}$, immediately satisfying condition \eqref{eq:KL}. For states within the DFS to be sensitive to $\omega_0$ we also need $\ket{0_\textsc{l}}$ and $\ket{1_\textsc{l}}$ to be non-degenerate energy eigenstates, so that $H P \not \propto P$, satisfying condition \eqref{eq:KL_H}. Such a DFS satisfies the ECQS conditions---accordingly, our discussion of general ECQS encompasses DFS-enhanced sensing as a special case.

DFS-enhanced sensing is only possible for a small family of correlation matrices $C$ which we discuss below. A code designed for some general $C$, in contrast, will usually necessitate an active recovery $\mathcal{R}$. Given a projector $P$ onto the code satisfying the ECQS conditions, an appropriate choice of $\mathcal{R}$ is the usual so-called transpose channel\cite{zhou:2017, nielsen:2000}. We summarize a standard way of implementing it in the Methods. For a logical state $\rho = P \rho P$ this recovery gives $\mathcal{RD}(\rho) = 0$ and $\mathcal{RH}(\rho) = [H_\text{eff}, \rho] \neq 0$ as desired, where $H_\text{eff} = P H_0 P$ (ref.\ \onlinecite{zhou:2017}). In other words, the sensor approaches noiseless evolution by $H_\text{eff}$ in the limit of frequent error detection/correction (i.e., $\Delta t \rightarrow 0$).

Conditions \eqref{eq:KL} and \eqref{eq:KL_H} seem to impose a stringent requirement on the $C$'s amenable to ECQS under signal and noise which are ``parallel", in that they couple to the sensor through the same operators (along the $z$ direction, i.e., the qubits' energy gaps, here). This need not be the case, however, since error correction can enhance quantum sensing even if it does not give a strictly noiseless limit. Notice in Fig.~\ref{fig:neg_sensitivity}, for instance, that ECQS enhances sensitivity for all $\gamma>0$, even though conditions \eqref{eq:KL} and \eqref{eq:KL_H} are only satisfied exactly when $\gamma=1$. (Another example is analyzed in the Supplementary Information, as is the robustness of our scheme.) More generally, if instead of satisfying \eqref{eq:KL} exactly, $P L_i^\dagger L_j P = m_{ij} P + O(\epsilon)$,  then $\mathcal{R D}(\rho) = O(\epsilon)$ for a logical $\rho$ instead of vanishing exactly \cite{zhou:2017}. If the time between successive recoveries is nonzero ($\Delta t > 0$) as in most experiments, then decoherence will appear in the logical dynamics at order $O(\Delta t/T_2)$ in $\mathcal{L}_\text{eff}$. (We write $O(t/T_2)$ rather than $O(||\mathcal{L}|| \, \Delta t)$ to simplify notation, assuming $\omega_0$ to be a small compared to $1/T_2$. If $\omega_0$ is not small, $O(\Delta t/T_2)$ should be taken to mean $O(||\mathcal{L}|| \, \Delta t)$.) Provided $\epsilon \ll \Delta t / T_2$, then, small violations of condition \eqref{eq:KL} will not appreciably change the degree to which quantum error correction suppresses noise in a sensor. This is true for generic ECQS schemes. When correcting noise which couples like the signal, in particular, allowing $\epsilon \neq 0$ enables codes which---by design---do not correct for errors $L_k$ with $||L_k|| \approx 0$, corresponding to an eigenvalue $\lambda_k$ of $C$ which is small but not exactly zero. Therefore in the present setting, relaxing condition \eqref{eq:KL} reduces the need for fine-tuned noise correlations.

For quantum error correction to filter noise from signal when $\Delta t / T_2$ is finite, $\mathcal{R}$ must suppress the former more than the latter. Choosing $\ket{0_\textsc{l}}$ and $\ket{1_\textsc{l}}$ to be eigenstates of $H_\text{eff}$ with $E_0 > E_1$, the effective Hamiltonian takes the form $H_\text{eff} =  \alpha P + \frac{\omega_\textsc{l}}{2} Z_\textsc{l}$, where $Z_\textsc{l} = \ket{0_\textsc{l}} \! \bra{0_\textsc{l}} - \ket{1_\textsc{l}} \! \bra{1_\textsc{l}}$ and $P$ acts as $I$ on the code. Just as $\epsilon$ and $\Delta t/T_2$ describe the extent to which noise is suppressed through frequent error correction, the ratio $A_\omega = \frac{\omega_\textsc{l}}{\omega_0}$ describes the signal gain; that is, the fraction of the physical signal that survives at the logical level. Together with previous arguments about \eqref{eq:KL}, we arrive at sufficient conditions in terms of this signal gain for error correction to enhance quantum sensing: $P L_i^\dagger L_j P = m_{ij} P + O(\epsilon)$, and $A_\omega \gg \epsilon$. There is an analogy with dynamical decoupling to be drawn here: both quantum error correction and DD can significantly enhance sensing by partially filtering noise from the signal---they need not remove the noise entirely to be useful.

This analogy goes further: Just as DD sequences must be tailored to sense in a particular frequency band of interest, error-correcting codes must be tailored to $C$ and $\vec{h}$ for a sensor to measure only in a particular spatial ``mode", in order to correct for noise which couples locally in the same way as the signal. That is, a particular $\delta \omega$ and arrangement of sensing qubits (likely determined at the time of fabrication) will require a unique $P$. This is because the scheme depends on a code \textit{not} correcting for $\vec{v}_k \cdot \vec{Z}$ with $\lambda_k \sim 0$, thus allowing the component of $H_0$ along $\vec{v}_k \cdot \vec{Z}$ to affect the logical states. Therefore, we expect that in experiment, codes will need to be tailored for individual devices, much like control sequences must be. We present here a robust method of doing so.

When $\vec{h} \notin \text{col}(C)$, refs.\ \onlinecite{demkowicz:2017, zhou:2017} provide recipes for codes which exactly satisfy the ECQS conditions, although these may require the sensor to contain up to $N$ noiseless ancilla qubits which do not couple to $\omega$, in addition to the $N$ sensing qubits. Here we take a different approach which naturally tolerates small violations of the ECQS conditions, such as those discussed above. That is, it allows one to find codes which enhance sensitivity, irrespective of whether they provide a noiseless limit in theory. In contrast with several previous works\cite{demkowicz:2017, zhou:2017}, our approach does not require the overhead of additional ancillas as part of the code. Specifically, for a given $C$, we map the task of finding a $P$ for \eqref{eq:KL} and \eqref{eq:KL_H} to an optimization problem, whose solutions are codes satisfying $P L_i ^\dagger L_j P = m_{ij} P + O(\epsilon)$ for some $M = M^\dagger$, and giving a minimum signal gain of $A_{\omega, \, \text{min}}$ that is freely adjustable:
\begin{gather} \label{eq:optimization}
\!\!\!\!\!\!\!\!
 \text{Minimize } F_\text{tot} 
= 
\sum_{E \in \mathcal{S}} F_E \\
\text{subject to } 
F_G >   A_{\omega, \, \text{min}}^2 
\text{ and } 
\braket{x}{y} = \delta_{xy}, \nonumber
\end{gather}
where $G := \frac{1}{2} \vec{h} \cdot \vec{Z} = H_0/\omega_0$ and 
\begin{equation}
F_E \big(\ket{x}, \ket{y} \big) = 
\big |
\bra{x} E \ket{x} - \bra{y} E \ket{y}
\big |^2
+
4 \big | \bra{x} E \ket{y} \big|^2. 
\label{eq:objective}
\end{equation}
Notice that $F_\text{tot}$ is non-negative with zeros where $P = \ket{x}\!\bra{x} + \ket{y}\!\bra{y}$ satisfies condition \eqref{eq:KL}. In fact, solutions to $F_\text{tot} = 0$ with $A_{\omega, \, \text{min}} = 0$ exactly satisfy the ECQS conditions and vice versa. Relaxing these conditions slightly, one can find codes approximately satisfying \eqref{eq:KL} and \eqref{eq:KL_H} by using $F_\text{tot} \le \epsilon^2$ as a convergence criterion and $A_{\omega, \, \text{min}} \gg \epsilon$. Note that the resulting codes can be quite general; for instance, they need not be stabilizer codes. Further details are provided in the Methods.

\begin{figure}[h]
\raggedleft
\includegraphics[trim=50pt 0pt 0pt 15pt, clip=true, width=0.45\textwidth]{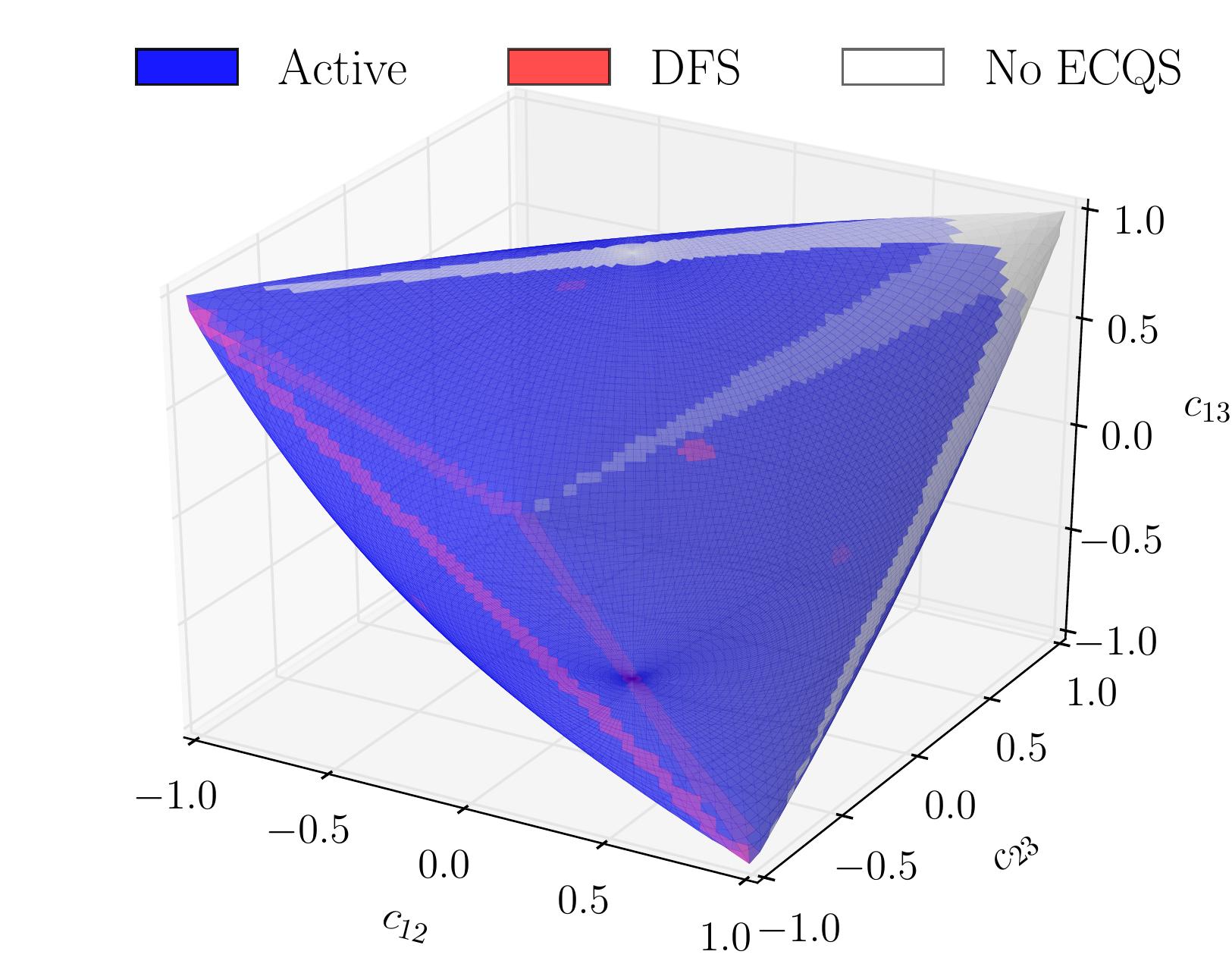}
\caption{Values of $c_{12}$, $c_{23}$ and $c_{13}$ for which there exists a three-qubit code satisfying the ECQS conditions to a tolerance of $\epsilon = 10^{-5}$ and $A_{\omega, \, \text{min}} = 10^{-1}$, and not requiring noiseless ancillas. The points $(c_{12}, c_{23}, c_{13})$ approximately satisfying conditions \eqref{eq:KL} and \eqref{eq:KL_H} form a tetrahedron-like surface. The portions of the surface in blue denote $C$'s for which ECQS is possible with an active (i.e., non-trivial) recovery. The red regions (enlarged for visibility) denote $C$'s for which DFS-enhanced sensing is possible, and the white regions denote $C$'s for which the optimization in Eq.~\eqref{eq:optimization} failed to converge to within the specified tolerance, either because the achievable signal gain is too small, or because of poor local minima in $F_\text{tot}$. The continuous red band comprises $C$'s for which noise on a pair of qubits is perfectly anti-correlated ($c_{ij}=-1$). Notice that ECQS is generically possible for both positive and negative noise correlations; it fails here only when $c_{ij} \approx 1$ for some pair of qubits ($i\neq j$), since this gives a missing/subdominant jump operator orthogonal to $H_0$.}
\label{fig:correlations}
\end{figure}

For a two-qubit sensor, $C$ has a zero eigenvalue only when $c_{12} = \pm 1$. The $c_{12} = 1$ case has $\vec{h} \in \text{col}(C)$ and is therefore not amenable to ECQS. The $c_{12} = -1$ case, on the other hand, has $\vec{h} \notin \text{col}(C)$. Therefore, $N=2$ sensing qubits under strongly anti-correlated noise ($c_{12} \approx -1$) can benefit from ECQS---in fact, they can be used for DFS-enhanced sensing. (Both $c_{12} \approx +1$ and $-1$, however, could be useful for gradiometry, i.e., to measure a mean difference between the energy gap of each qubit.) For a three-qubit sensor a much broader family of $C$'s can satisfy the ECQS conditions. Using the mapping described above, the $C$'s for which Eq.~\eqref{eq:optimization} yielded codes approximately satisfying conditions \eqref{eq:KL} and \eqref{eq:KL_H} with no ancillas are shown in Fig.~\ref{fig:correlations}. Notice that DFS-enhanced sensing with $N=3$ qubits is only possible for a small family $C$'s. Therefore, while such schemes may be powerful \cite{roos:2006, chwalla:2007, dorner:2012, jeske:2014}, it could be exceedingly difficult to engineer the spatial noise correlations they require in many devices. Codes with active recoveries, in contrast, are much more broadly applicable. A scheme to measure $C$ in experiments is given in the Methods.

We have assumed for simplicity so far that the coupling strength of each qubit to the external field $\omega$, parameterized by $h_i = 1$, is the same for all qubits $i$. This assumption is of course not necessary; the results presented here hold (with trivial adjustments) for arbitrary values $h_i' \neq 0$ which may vary across qubits. Moreover, whether or not ECQS conditions can be satisfied depends only on the spatial profile of $\delta \omega$ and on the locations of the qubits, not on the coupling strengths $\vec{h}'$. We reserve the proof for the Methods section.

\section*{Discussion}

We have shown how error-corrected quantum sensing can filter noise from a signal when both couple to a sensor locally through the same operators. This stands in contrast with earlier explicit ECQS schemes, which have been limited to correcting noise separate from the quantity to be measured, in that it couples differently to the sensor. In many quantum sensors such noise is sub-dominant, while the type of noise considered here is the limiting source of decoherence, and can only be partially filtered through DD. Our scheme relies on the observation that Eqs.~\eqref{eq:KL} and \eqref{eq:KL_H} can be viewed as a condition on the spatial correlations of the signal and noise. This view raises a close parallel between ECQS and DD: for signal and noise which couple identically to a sensor (in the sense of Eq.~\eqref{eq:H_stochastic}), ECQS and DD can enhance sensitivity by acting as filters in the spatial and frequency domains, respectively. However, since these two schemes separate noise from signal on totally separate grounds, they are complementary, in that the limitations of one are not shared by the other. Finally, we proposed a numerical method of tailoring ECQS codes to the noise observed in specific devices. It yields not just codes that correct noise perfectly in the $\Delta t \rightarrow 0$ limit, but also codes which can more generally improve sensitivity. We applied this method to sensors comprising 2 and 3 qubits---with no extra ancillas in the code---and showed how our error correction scheme could provide an advantage even in relatively small devices.

Both ECQS and DD filter noise which couples locally like the signal by exploiting correlations in it: spatial correlations in the case of ECQS, and temporal correlations for DD. Accordingly, the effectiveness of both schemes depends on the degree to which noise in a sensor can made to have suitable correlations. (Note that different spatial noise correlations may lend themselves best to different sensing tasks. For instance, uniform positive correlations yield a dominant noise mode proportional to $H_0$. This makes them them ill-suited for measuring small field values through ECQS, but well-suited for gradiometry.) Engineering appropriate spatial noise correlations is likely to be highly implementation-dependent, as it is with temporal correlations\cite{ajoy:2016}. This is because the main sources of noise can be entirely different in different types of quantum sensors. While an analysis of achievable noise correlations in various types of sensors is beyond the scope of the present work, we note here simply that strong spatial correlations have been reported already in several experiments, e.g., refs.\ \onlinecite{fortunato:2002, roos:2006, chwalla:2007,  monz:2011, schindler:2011, romach:2015}.

The scheme presented here exploits spatial correlations to extract signal from a noisy background, which generically causes dephasing. (Of course, the signal and noise in question need not couple to the sensor via $\sigma_z$; in general, the qubits could be damped along any axis.) A similar approach may be possible with generalized amplitude-damping ($T_1$-type) errors, which are second only to phase errors as the dominant decoherence mode in many quantum sensors. That is, a sensor with qubits made to thermalize collectively, rather than individually, could be amenable to quantum error correction. The approach presented here could be combined with such a scheme---or with previous ECQS schemes---raising the intriguing prospect of a quantum sensor that is error-corrected against noise in all three spatial directions. While correlated errors can often be detrimental to error-corrected quantum computation (see, e.g., ref. \onlinecite{aharonov:2006}), they may prove a valuable resource for error-corrected quantum sensing.

\section*{Methods}

\subsection*{Proof: Equivalence of Eq.~\eqref{eq:hcolC} and ECQS conditions}

We show here that Eq.~\eqref{eq:hcolC} is equivalent to $H_0 \notin \mathcal{S}$ for the background noise described in Eq.~\eqref{eq:Lindblad}, where $\mathcal{S}$ is the Lindblad span. Let $\{\vec{v}_i \} \subset \mathbb{R}^N$ be an orthonormal eigenbasis of $C$ such that $C \vec{v}_i = \lambda_i \vec{v}_i$, and define $\tilde{L}_i := \vec{v}_i \cdot \vec{Z} = \tilde{L}_i^\dagger$ and $L_i := \sqrt{\lambda_i} \tilde{L}_i = L_i^\dagger$. Notice that $\langle \tilde{L}_i, I \rangle = \langle \tilde{L}_i, \tilde{L}_j \tilde{L}_\ell \rangle = 0$  under $\langle A, B \rangle = \tr(A^\dagger B)$, so $\mathcal{S}$ can be decomposed into orthogonal subspaces as $\mathcal{S} = \mathcal{S}_1 \oplus \mathcal{S}_2$ where $\mathcal{S}_1 := \text{span} \{ L_i \}_{i\ge 1}$ and $\mathcal{S}_2 := \text{span} \{ I, L_i L_j \}_{i,j \ge 1}$. Having diagonalized $C$, we can express $H_0$ in terms of $\tilde{L}_i$'s (rather than $Z_i$'s) as $H_0 = \sum_{i=1}^N \alpha_i \tilde{L}_i$ for unique coefficients $\alpha_i = 2^{-N} \tr( \tilde{L}_i H_0 ) = \frac{\omega_0}{2} \, \vec{v}_i \cdot \vec{h}$, implying that $H_0 \perp \mathcal{S}_2$. Therefore, $H_0 \notin \mathcal{S}$ if and only if (iff) $H_0 \notin \mathcal{S}_1$. This happens iff there is a $k$ such that $\lambda_k=L_k=0$ and $\alpha_k \neq 0$, or equivalently, iff $\vec{v}_k \cdot \vec{h} \neq 0$ for some $\vec{v_k} \in \text{ker}(C)$. Finally, since $C = C^\top$ we have $\text{col}(C) \oplus \text{ker}(C) = \mathbb{R}^N$, and so $H_0 \notin \mathcal{S}$ iff $ \vec{h} \notin \text{col}(C)$.

\subsection*{Recovery Channel}

We describe here a standard way of implementing the so-called transpose recovery channel, following refs.\ \onlinecite{zhou:2017, nielsen:2000}. Given a known $P$ such that $P L_i^\dagger L_j P = m_{ij} P$, we have $P (L_i - m_{0i} I)^\dagger (L_j - m_{0j} I)P = \tilde{m}_{ij} P$ for some Hermitian $\tilde{M} = (\tilde{m}_{ij})_{i,j\ge 1}$. Let $W$ be a unitary matrix such that $W^\dagger \tilde{M} W = \text{diag}(d_1, d_2, \dots)$, and $E_i := \sum_{j} w_{ji} (L_j - m_{j0} I)$. For $d_i \neq 0$ one can find a unique unitary $U_i$ such that $E_i P = \sqrt{d_i} U_i P$ via polar decomposition. The channel $\mathcal{R}$ can then be implemented by performing a projective measurement in $\{P_0, P_1, P_2, \dots\}$, where $P_i := U_i P U_i^\dagger$ and $P_0 := P$, then applying $U_i^\dagger$ for outcome $i$, where $U_0 := I$. (N.b., if $\text{rank}(\tilde{M}) = 1$ this procedure is trivial, so the code forms a DFS.) 

\subsection*{Numerical Code Search}

The objective function $F_\text{tot}$ may have several distinct zeros satisfying the constraints with $A_{\omega, \, \text{min}}=0$; for instance, the logical states in Eq.~\eqref{eq:ex1_states} and \eqref{eq:ex2_states}. (In other words, there can be more than one $P$ exactly satisfying Eqs.~\eqref{eq:KL} and \eqref{eq:KL_H}.) On the other hand, it will have no such zeros when $\vec{h} \in \text{col}(C)$. More generally, for given $\epsilon, A_{\omega, \, \text{min}} \ge 0$ there may exist multiple regions where $F_\text{tot} \le \epsilon^2$ subject to the constraints, or there may exist none for $C$ not amenable to ECQS with $N$ sensing qubits.

A similar approach to finding codes was recently used in ref.\ \onlinecite{albert:2017}, although to our knowledge it has not previously been used for ECQS. The factor of 4 in Eq.~\eqref{eq:objective} was included specifically for the purpose of finding codes for sensing: While this factor is irrelevant for enforcing that $E \in \mathcal{S}$ be corrected, a simple calculation shows that $F_G$ for a code gives exactly its signal gain squared. Therefore, requiring that $F_G > A_{\omega, \, \text{min}}^2$ and $F_\text{tot} \le \epsilon^2$ for some $A_{\omega, \, \text{min}} \gg \epsilon$ is a transparent way of demanding that quantum error correction suppress noise much more strongly than the signal.

\subsection*{Measuring $C$}

In an experiment, finding an appropriate code for ECQS first requires knowledge of the noise correlations encoded in $C$. For qubits $i$ and $j$, the coefficient $c_{ij}$ can be inferred by preparing the GHZ state $\frac{1}{\sqrt{2}}(\ket{0_i}\ket{0_j} + \ket{1_i}\ket{1_j})$ and measuring its pure dephasing rate $\Gamma_{ij}$, which is related to $c_{ij}$ through $\Gamma_{ij} = \frac{2}{T_2} (1 + c_{ij})$, after subtracting the dephasing due to any relaxation that might also be present in practice. Note that while we arrived at Eq.~\eqref{eq:Lindblad} by considering a signal with a noisy background, the physical source of dephasing is largely immaterial; all that matters is its spatial correlation profile $C$.

\subsection*{Proof: Independence from Coupling Strengths}

We show here that whether Eq.~\eqref{eq:hcolC}, and therefore also the ECQS conditions, is satisfied does not depend on the coupling strength of each qubit to the external field. Defining $D = \text{diag} (\vec{h}')$, the Lindblad equation for general coupling strengths $\vec{h}'$ has $H_0' = \frac{\omega_0}{2} \vec{h}' \cdot \vec{Z}$ and $C' = D C D$. Observe that $\vec{h}' = D \vec{h}$ and $\text{ker}(C') = \{ D^{-1} \vec{x} \, | \, \vec{x} \in \text{ker}(C) \}$, so $\vec{h'} \notin \text{col}(C')$ if and only if $\vec{h} \notin \text{col}(C)$. 

\subsection*{Code availability}

The optimization in Eq.~\eqref{eq:optimization} was performed using the SciPy Python library (v0.17). We employed a basin-hopping global optimization algorithm, which ran a sequential least squares programming (SLSQP) local optimization at each step. The full computer code used to generate these results is available from the corresponding author upon request.

\subsection*{Data availability}

The numerical datasets generated during and/or analysed during the current study are available from the corresponding author on reasonable request.\\

Supplementary information is available at the npj Quantum Information website.

\section*{Acknowledgements} 

We wish to thank C\'edric B\'eny, Liang Jiang, and Sisi Zhou for insightful discussions.

\section*{Competing Interests}

The authors declare no competing financial interests.

\section*{Contributions}

D.L.\ and P.C.\ conceived the novel QEC scheme for sensing. D.L.\ developed the theory and performed the computations. All authors discussed the results and contributed to the final manuscript.

\section*{Funding}

This work was supported in part by the U.S. Army Research Office through grant No. W911NF-15-1-0548, by the NSF PHY0551153 and 1641064 grants, and by the NSERC PGS-D program.

\end{document}